# Liquid Metal Enabled Electrobiology: A Generalized Easy Going Way to Tackle Disease Challenges

Xuelin Wang[1*], Yi Ren[1*], Jing Liu[1,2]**

1. Department of Biomedical Engineering, School of Medicine,

Tsinghua University, Beijing 100084, China

2. Beijing Key Lab of CryoBiomedical Engineering and Key Lab of Cryogenics,

Technical Institute of Physics and Chemistry,

Chinese Academy of Sciences, Beijing 100190, China

[*]X. Wang and Y. Ren contributed equally to this work

**Address for correspondence:

Dr. Jing Liu

Department of Biomedical Engineering,

School of Medicine,

Tsinghua University,

Beijing 100084, China

E-mail address: jliubme@tsinghua.edu.cn

Tel. +86-10-62794896

Fax: +86-10-82543767

**Abstract:**

In this article, a new conceptual biomedical engineering strategy to tackle modern disease challenges, termed as liquid metal enabled electrobiology, is proposed. This generalized and easy going way is based on the physiological fact that specially administrated electricity would induce a series of subsequent desired biological effects, either shortly, transitional or permanently. Owing high compliance within any part of the biological tissues, the liquid metal would aid to mold a pervasive way to treat physiological or psychological diseases. As highly conductive and non-toxic multifunctional flexible materials, such liquid metals (LMs) consist of the core to generate any requested electric treating fields (ETFields) which can adapt to various sites inside the human body. The basic mechanisms of electrobiology in delivering electricity to the target tissues and then inducing expected outputs for disease treatment are interpreted. The methods toward realizing soft and conformable electronics based on liquid metal are illustrated. Further, a group of typical disease challenges were taken to illustrate the basic strategies to perform liquid metal electrobiology therapy, which include but are not limited to: tissue electronics, brain disorder, immunotherapy, neural functional recovery, muscle stimulation, skin rejuvenation, cosmetology and dieting, artificial organs, cardiac pacing and cancer therapy etc. Some practical issues involved in the electrobiology for future disease therapy were discussed. Perspective along this direction to incubate an easy-going biomedical tool for health care was pointed out.

# 1. Introduction

In the cutting-edge field of disease therapy, physiotherapy has received great attention by means of some physical agents such as electricity, magnetism, sound, light, heat, freezing, water, mechanical force etc. Typical physiotherapy methods generally include: electrotherapy, magnetotherapy, ultrasound therapy, phototherapy, hyperthermia and cryotherapy [1-3]. Compared to the traditional disease treatments like chemotherapy, radiotherapy and surgical resection, physiotherapy can prevent and treat diseases and restore or reconstruct body function with positive efficacy, minimal invasion and high survival rate. Particularly, electrotherapy has displayed excellent advantages in disease therapy, in some cases, preventing or treating cancer [4], via regulating cell apoptosis or proliferation [5].

Electrotherapy, using electrical energy as medical tool, has been applied for disease treatment since 1855 and has offered significant therapeutic effect [6]. In principle, electrobiology refers to a series of numerous physiological effects caused due to electrical stimulating as human body is composed with plenty of water and conductive electrolyte. This consists of the fundamental of electrotherapy. For example, the activation and inhabitation of cells, like muscles and cardiac cells, can be controlled by resting potential and action potential [7, 8]. Electrical signal is also an important part in neuron communication [9]. What's more, electricity can affect the growth and differentiation of cells and voltage-gated ion channel is important in substance transportation [10]. Electrical signal can even be found during photosynthetic process of Chloroplast [11]. Since electrical signal plays such an important role in physiological including psychological events, external electrostimulation (ES) can be applied to influence nearly all the states of cells and tissues and restore some missing functions of injured organs. There have already been various applications of electrotherapy, including pain management, neuromuscular dysfunction, joint mobility, tissue repair, and acute or chronic edema [6]. However, the development of electrotherapy comes across some difficulties too. In order to guarantee the best physiological effects, it is essential that appropriate energy is absorbed by the particular tissue. But common rigid electrodes generally show less flexibility and biocompatibility, which limits the applications of electrotherapy. Thus, a customizable ES system depending on different parts of body is rather necessary. On the whole, a novel flexible material − room temperature liquid metal (RTLM), which can match the body disease models perfectly, presents more and more advantages in electrotherapy, especially for the targets with abnormal shape, such as tumors, blood vessels, bones, and cavities etc [12].

Generally speaking, RTLM refers to the liquid metal (LM) whose melting point is around room temperature, such as gallium, bismuth, lead, tin, cadmium, indium and its alloys [13]. Thereinto, gallium and its alloys are often applied with different composition [14]. For example, $GaIn_{24.5}$, the most common LM, refers to the mixture of 75.5% gallium and 24.5% indium by weight. However, mercury, a kind of traditional RTLM, is excluded from ES material on account of its high toxicity to human body. RTLM offers the unique natures of metallicity and fluidity to manufacture flexible electronics [14, 15], which makes it especially suitable for different electric field shapes on soft human tissues. Due to the high electrical conductivity of LM $GaIn_{24.5}$ alloy ($3.4\times 10^6$ S/m) compared with conductive carbon ($1.8\times 10^3$ S/m) and conductive polymer PEDOT:PSS ($8.25\times 10^3$ S/m) [16], its ability to conduct electricity is much better than that of non-metal. In addition, when bending and twisting LM printed wire, the resistance variation is not obvious, which manifests its reliability in soft printed electronics [17-20]. Furthermore, LM has excellent wettability on different substrate materials with different surface roughness and material properties [15, 21]. More importantly, owing to the merits of non-toxicity and benign biocompatibility, such metal has made remarkable progresses in bone cement replacement [22], high-resolution angiography contrast [23], nerve connection [24, 25], human exoskeleton [26], drug delivery nanomedicine [27] etc. Overall, the remarkable features of RTLM, such as high electrical conductivity, excellent flexibility, good wettability and fine biocompatibility indicate its wide usage in electrobiology for disease therapy [28].

In this perspective article, we are dedicated to present a comprehensive interpretation on the new conceptual liquid metal electrobiology. A series of potential ways for disease therapy based on the biological effects induced by electricity will be outlined. To clarify the foundation of LM to enable electrobiology on human cell, several basic mechanisms are mainly discussed. Finally, we provide an overview on the practical issues of LM enabled electrobiology, and highlight prospects for further research.

## 2. Foundation of LM Enabled Electrobiology

**2.1 Flexibility and conformability of LM bio-electrode**

Form Figure 1A, one can know that LM has the lowest Young's modulus value than that of rigid materials (polyimide, copper, graphene etc.), elastomers (PDMS (polydimethylsiloxane), skin etc.) and even gels (brain tissue, muscle etc.) [29]. The most flexible material can adapt to complex tissue shapes to achieve any desired biological performance and thus be prepared to various conformable bio-electrodes. Figure 1B depicts a conceptual illustration of LM soft electrode on body cavity. Clearly, the flexibility of LM well avoids damaging human tissue when sued as electrobiology bio-electrode inside human body [12]. Meanwhile, LM can be shaped to arbitrary electrode shapes through certain molds (Figure 1C) [12]. Due to its conformability to any complex surface, LM soft electrode has been introduced to tumor therapy (Figure 1D) [30]. Besides, LM can be conformably attached on skin surface without causing any gap which would significantly reduce the contact resistance between skin and electrode (Figure 1E) and thus provide better signal stability and reduction of noise than conventional Ag/AgCl electrodes [31]. The presentation of numerous LM bio-electrode suggests that LM owns outstanding reliable flexibility and conformability when employed in LM enabled electrobiology.

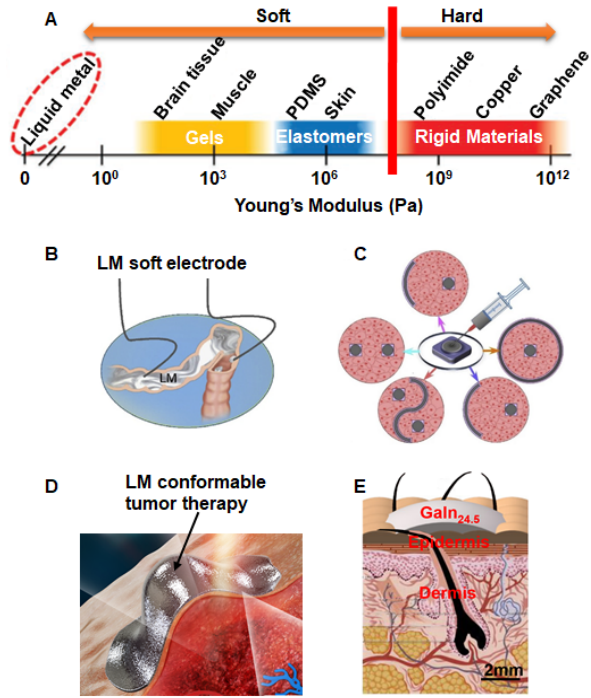

**Figure 1.** Various flexible and conformable LM bio-electrodes. (A) The comparison of Young's modulus among liquid metal and common soft and rigid materials [29]. (B) The conceptual illustration of LM soft electrode [12]. (C) LM conformable electrode with five different shapes. (D) LM conformable tumor therapy [30]. (E) Structural diagram of LM soft electrode on skin surface [31].

## 2.2 Electrobiology in cell living/dying

Research has found that electricity can induce cell apoptosis or proliferation, closely depending on the voltage amplitude and frequency. For instance, electrostimulation greater than 5V/mm can change the permeability of cell membranes, which increases the rate of cell apoptosis [32-36], while electrostimulation less than 5V/mm can promote cell proliferation mainly by upregulating growth factors (GF), activating proliferation signaling pathways, increasing intracellular $Ca^{2+}$ and interfering with the cell cycle [37-42]. What's more, through applying direct current electric fields, cell migration can be induced, which is important in tissue formation, organ regeneration and wound healing [43-46]. Different factors and signal pathways are involved during these changes. A more detailed description of these mechanisms is displayed in Table 1.

Exposing normal or cancer cells to alternating electric fields (1-3 V/cm, 100-300 kHz), dividing cells can be influenced while quiescent cells cannot (Table 1) [47, 48]. All charges and polar molecules are forced to alternate direction, which can disrupt the separation of chromosomes (Fig. 2A). Traction can be calculated by Bahaj and Bailey theory [49], i.e.

$$F_{DEP} \propto \frac{V^2}{L_e^3} \qquad (4)$$

where, $F_{DEP}$ is the dielectric force. $V$ is the applied voltage. $L_e$ is the distance between electrodes. The formation of internal structure is damaged, which interferes with cell mitosis and ultimately destroys the cell. LM can be utilized to produce conformable ETFields to enhance the therapeutic effect in electrobiology therapy.

Table 1. Effect of electrobiology on cells and related mechanism.

| Effect | Electrostimulation | Mechanism | Ref. |
| --- | --- | --- | --- |
| Disrupt cell mitosis | 1-3 V/cm, 100-300 kHz) | (1) In mitosis cells, the non-uniform distribution of electric field induces high electric-field intensity in the junction of two daughter cells (Fig. 2A). All charges and polar molecules are forced to alternate direction, which can disrupt the separation of chromosomes.<br>(2) Because of the uniform distribution of electric field in quiescent cells, there is almost no damage to normal cells. | 47, 48 |
| Increase cell apoptosis | Greater than 5V/mm | (1) ES makes excessive $Ca^{2+}$ influx due to the pore formation/electroporation in cell membrane.<br>(2) Caspase 3, a key player in apoptosis, is activated.<br>(3) The damaged DNA induces MEK1/2 phosphorylation. Then the activated ERK1/2 by phosphorylated MEK1/2 prevents cell proliferation by sequestering ERK1/2 in cytosol, which can increase cell cycle progression from G to S phase | 5, 32-36 |
| Increase cell proliferation | Less than 5V/mm | (1) ES increases the secretion of growth factors.<br>(2) Akt prevents cell apoptosis by inhibiting the pro-apoptotic factors Bim, caspase 9, Bax, Bad and FOXO-3. And Akt promotes the cell survival by activing transcription factor NF-KB to advance transcription of pro-survival genes.<br>(3) ERK1/2 is translocated to the nucleus to activate G to S phase transition. | 5, 37-42 |
| Activate cell migration | Migration velocity depends on cell type and voltage amplitude | (1) NaKA and NHE3, considered as a switch of cathodal/anodal migration, accumulates at the cathodal or anodal edge of the cells, causing cell depolarization and cytoskeleton redistribution.<br>(2) PI3K, which is found to accumulate and activate at the leading edge of cells with downstream effectors, induces cell cathodal migration.<br>(3) PTEN is more likely to activate at the anodal side rather than cathodal side. | 43-46 |

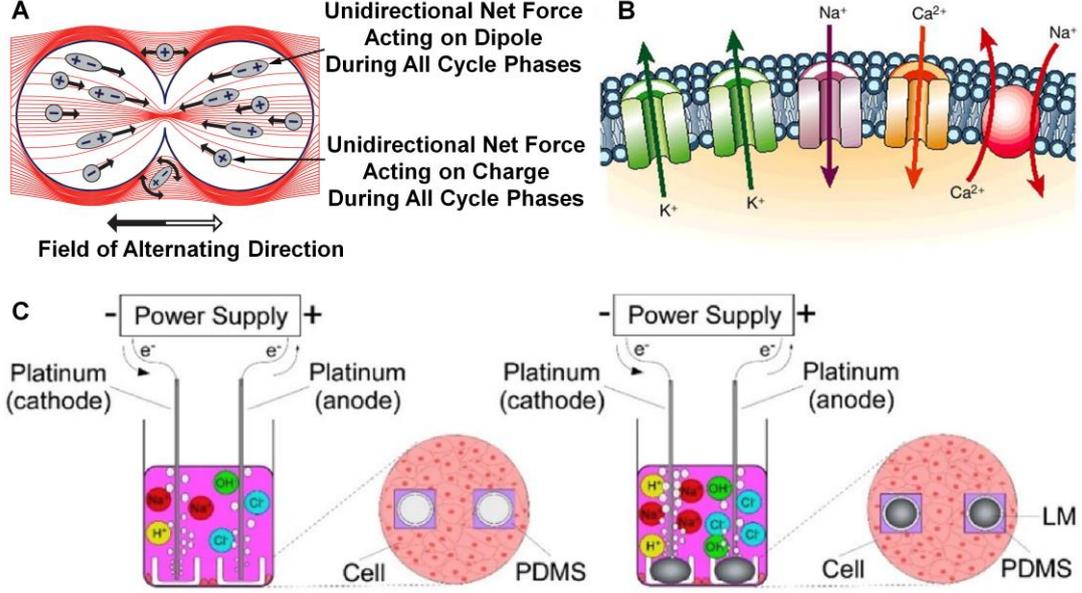

**Figure 2.** (A) The effect of electrical field treatment on dividing cells [48]. (B) The ion channels of $K^+$, $Na^+$, $Ca^{2+}$ and electrogenic transporters in cardiac cells [51]. (C) LM electrode on tumor cell therapy in vitro under electricity [12].

All living cells have transmembrane electrical potential difference and the change of electrical potential controls the opening or closing of ion channels. These voltage-gated ion channels can regulate ion movement across cell membrane to reach a balance [50], and Fig 2B is the voltage-gated ion channel in cardiac cells [51]. When the system is at thermodynamic equilibrium, we can get the transmembrane voltage V by Boltzmann distribution [52], i.e.

$$\frac{P_i}{P_j} = \exp\left(-\frac{U_i - U_j}{RT}\right) \quad (1)$$

where, $P_i$ and $P_j$ are the probability of state $i$ or $j$, respectively. $RT$ is a constant of the distribution. $U_i$ and $U_j$ are the state $i$ or $j$ energy, respectively. According to Formula 1, we know the relative energy of an ion $X$ on the inside or outside of membrane [32] which reads as

$$\frac{[X]_o}{[X]_i} = \exp\left(-\frac{U_o - U_i}{RT}\right) \quad (2)$$

When $I = 0$,

$$V = V_X = \frac{RT}{F} \ln \frac{P_K[K]_o + P_{Na}[Na]_o + P_{Cl}[Cl]_i}{P_K[K]_i + P_{Na}[Na]_i + P_{Cl}[Cl]_o} \quad (3)$$

where, $F$ is the Faraday's constant (96480 C/mol). $[X]_i$ is the concentration of ion $X$ on the inside of membrane. $[X]_o$ is the concentration of ion $X$ on the outside of membrane. When applying electricity on a cell, the resting transmembrane potential can be changed according to Equation 3, which induces cell apoptosis or proliferation [37, 53-55]. For instance, by inhibiting proliferation and inducing apoptosis, low voltage electric pulse kills the human squamous cell carcinoma cell lines [55]. Furthermore, LM electrode show better cell destruction effect than conventional platinum electrode under electricity with electrochemical effect on tumor cell therapy in vitro ((Fig. 2C) [12]. Hence, there are many applications of electrobiology in disease therapy, such as cancer therapy [32, 33], skeletal muscle atrophy [56], nerve defects [57] and wound healing [37] etc. The LM enabled electrobiology therapy on human with high targeting and low side effect changes the cell viability and cell cycle, promotes or inhibits tissue regeneration, thus realizing functional electrobiology disease treatment [58].

## 3. Basic Therapeutic Strategies of LM Electrobiology

The flexibility and electrical conductivity of RTLM makes it play an important role in electrobiology therapy. In fact, it is becoming a gradually used bio-material in implantable devices, electrical skin and wearable bioelectronics for its non-toxicity and benign biocompatibility [59-62]. As a kind of new fashioned flexible electronics, LM has been widely utilized in skin electronics for health monitoring and detection. Besides skin electronics, LM shows great potential in electrobiology, including brain stimulation, use as neural junction and muscle stimulation, cosmetology and dieting, artificial organs, cardiac pacing, cancer therapy, immunization therapy even psychotherapy etc. Fig 3 presents the typical applications and perspectives of LM enabled electrobiology, such as artificial retina [63], cochlear implant [64], flexible sensors [65], energy harvesting during movement [66], electrical skin [60, 67], electrical muscle stimulation [68], liquid metal bath electrode [69], nerve connection [59, 70], conformable tumor treatment, tumor treating fields [11, 71], pace-making and defibrillation, liquid metal electrode array [72] etc. Both experimental studies and clinical trials will enlarge its application scopes of LM enabled electrobiology therapy.

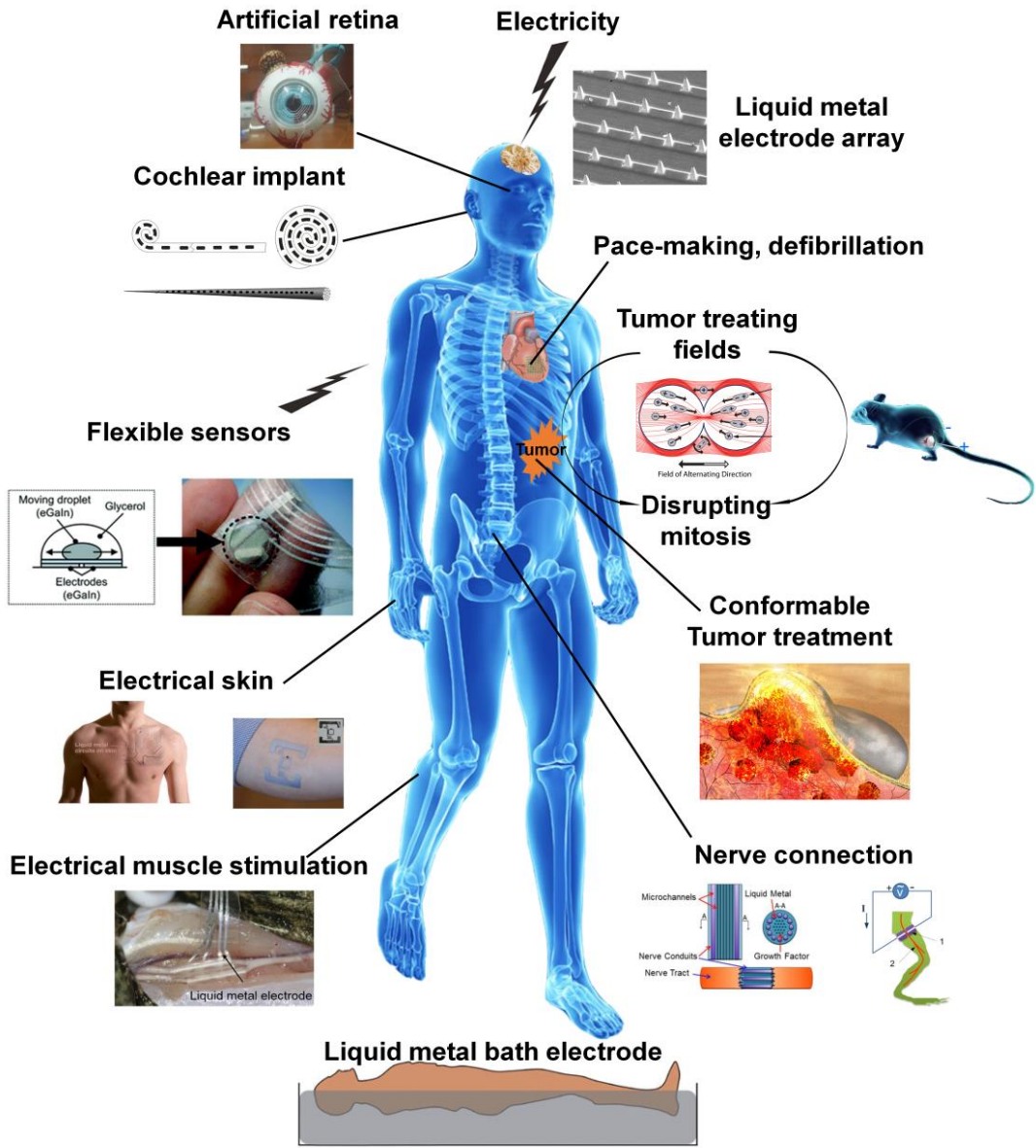

**Figure 3.** LM enabled electrobiology in disease therapy. Human model (http://cn.bing.com/); Artificial retina [63]; Cochlear implant [64]; Flexible sensors [65]; Electrical skin [60, 67]; Electrical muscle stimulation [68]; Liquid metal bath electrode [69]; Nerve connection [59, 70]; Conformable tumor treatment; Tumor treating fields [11, 71]; Pace-making and defibrillation; liquid metal electrode array [72].

### 3.1 Brain stimulation

Electrical brain stimulation (EBS) is to stimulate neurons or neural networks in the brain through direct or indirect excitation by electric current, usually for research or therapeutic purposes. If stimulating different parts of human brain, some acute effects could be induced, including sensory, motor, and autonomic, emotional or cognitive actions [73]. A typical application of EBS is deep brain stimulation, which provides hope to tackle Parkinson's disease [74], Alzheimer's disease [75] and major depression and obsessive-compulsive disorder [76]. What's more, the current pulse is applied to damaged nerves to promote nerve regeneration and restore cortical control of functional movement in a human with quadriplegia [77].

Electrodes for brain stimulation include surface electrodes and implanted electrodes. Wearable brain cap is a typical surface electrodes used in EEG recoding. To achieve more detailed EEG signal and better stimulation effects, implanted electrodes are applied rather than surface electrodes, especially for deep brain stimulation. Conventional implanted bio-electrodes generally require surgery which often causes severe side-effect to patients (Fig. 4A) [59]. Therefore, LM is applied to minimally invasive implantable biomedical devices through sequential injections of biocompatible packaging gelation and liquid metal ink (Fig. 4B). Because of the solidity of gelatin, various LM-based 3D flexible biological electrodes can be prepared to adapt to different requirements [50]. In addition, Liu et al have manufactured 3D microneedle electrode arrays to acquire electrical signals [72]. The SEM image in Fig 4C indicates its good consistency which reveals the potential for wider brain electrostimulation application. These LM-based implantable 3D bio-electrodes fabrication can be utilized in the manufacturing process of novel electrode cap for deep brain stimulation with noninvasive and precise feature.

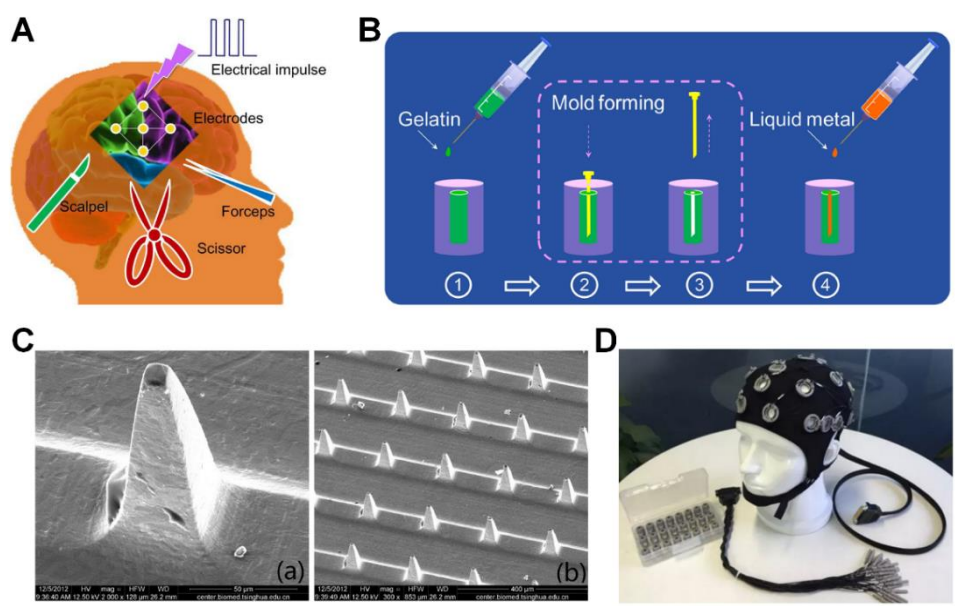

**Figure 4.** (A) Schematics for conventional surgery in implanting electrode for brain stimulator [59]; (B) The fabrication procedure of injectable metal electrodes [59]; (C) The SEM image of liquid metal electrode array [72]; (D) Brain electrode cap (http://cn.bing.com/).

**3.2 Muscle stimulation and neural functional recovery**

Electrical muscle stimulation is used as a training, therapeutic or cosmetic tool, including relaxation of muscle spasms, prevention or retardation of disuse atrophy, increasing local blood circulation, muscle re-education and maintaining or increasing range of motion [6]. Particularly, functional electrical stimulation can be used to generate muscle contraction in patients who have been paralyzed due to injuries of central nervous system. It has been proven that functional electrical stimulation can help prevent muscle atrophy after major knee ligament surgery [78]. Research on disease prevention through electrostimulation has been conducted and it has been found that electrical stimulation is beneficial to preventing type 2 fiber atrophy in early severe critical illness, and female pelvic floor dysfunction [79, 80]. Combining the function of high electric conductivity compared to other soft conductive materials with weak ability of stretching and folding, LM soft electronics designed for operation on the trunk or limbs of the body are attractive for electrical muscle stimulation [60, 67, 72, 81].

Most recently, a kind of LM electrode is developed, which is useful in neuromuscular stimulation [68]. As shown in Fig. 5D, LM is attached to excellently flexible PDMS substrate. The implanted LM electrodes are applied to sural nerve and tibial nerve of a dead bullfrog. The frog leg would lift up and down by stimulating sural nerve and tibial nerve alternately, which proves successful neuromuscular stimulation.

Surprisingly, it has also been demonstrated that LM can reconnect broken nerves and restore their function [70], which means neuromuscular electrical stimulation can be applied to not only complete nerves, but also broken nerves. As a kind of flexible implant in the neural tissues, LM is applied to connect nerves to achieve the long-term performance of implantable neuroprosthesis and neuronal electrical stimulation. Via galinstan, the present lab had recovered the functions of the peripheral neurotmesis for the first time, and a novel nerve junction material comes into being for patients to regenerate neural networks (Fig. 5C) [70]. The authors built up the bullfrog gastrocnemius model and introduced LM to link the snipped neural tissue. With a weak electrical stimulation, they proved the advantage of LM in nerve conduction. The result shows that the neural model connected by LM has a good stimulation signal which maintains the same consistency and fidelity as the normal neural tissue before cut. Due to its strong contrast under X-ray, LM can be easily taken out of the body after completing nerve repair to avoid secondary surgery. This method opens a new direction for neural connections and restoration.

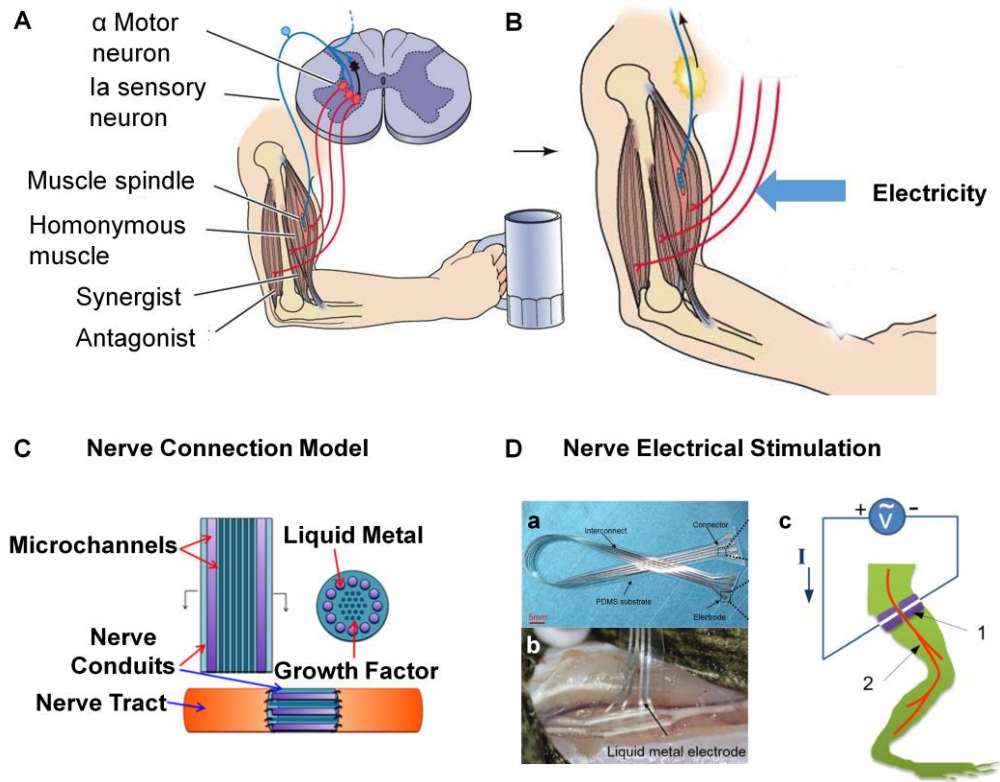

**Figure 5.** Nerve connection and implanted electrical stimulation. (A) and (B) respectively show excitation from central neuron and electricity stimulation to muscle (from NEUROSCIENCE, Fourth Edition, Fig 16.10). (C) Nerve connection model with liquid metal microchannels to repair the injured peripheral nerve [70]; (D) Implanted nerve electrical stimulation based on LM; (a) The optical image of LM electrode array [68]; (b) The implanted LM electrodes for the bullfrog sural nerve and tibial nerve [68]. (c) Schematic of nerve electrical stimulations based on the injectable LM electrode [59].

**3.3 Beauty and weight loss**

Skin care and weight loss are more and more popular nowadays. As the visible organ with multiple physiological functions, skin has been a hot topic in youthful study [82]. And obesity is tightly related to both beauty and health, thus attracting increasing attention. It has been found that electrical stimulation has promising potential in beauty therapy [83].

Painting LM on skin surface, electrical stimulation induces cell proliferation to make skin a better protective barrier that mitigates exposure to external factors such as extreme temperature, microorganisms, and radiation. Consequently, the youthful and healthful skin can appear via electrostimulation treatment. What's more, stimulating facial muscles and skin can improve circulation, activate fibroblasts and tone flaccid and sagging face muscle [83]. Conformable LM can even stimulate facial skin for better elasticity and smoothness.

Brown adipose tissue (BAT) is a heating-producing organ that plays an important role in decomposing human white adipose tissue (WAT) to reduce the risk of obesity [84-87]. Besides exercising, dieting or medicine regulating [88], electrical stimulation has been applied to active BAT thermogenesis with microcurrent on the dorsal surface of the tissue [89-93]. Due to the low-fidelity coupling between the traditional rigid metallic electrodes and dorsal skin surface [94, 95], RTLM is expected to be painted on the skin to reduce the energy loss on electrostimulation obesity.

It is also believed that electrical stimulation with wearable LM-based electronics makes the users more likely to participate in exercises as ES devices are comfortable and effective when doing sports. It allows the body to activate muscles without strain or stress and reduces the risk of injuries [83]. Combining LM-based wearable electrostimulation devices with proper exercises, the method of this healthy weight loss can do something for weight problems.

**3.4 Artificial retina and cochlear implantation**

Amit et al. have realized a soft coil based on LM and demonstrated that this coil is an excellent alternative to stiff and uncomfortable coils for biomedical implants [63]. LM was injected into elastomeric microfluidic channels to produce LM-based coils, which can deliver power to the implant coil efficiently in a telemetry system. Meanwhile, Zhao et al. have designed an innovative LM electronic coil for eye movement tracking (Fig. 6A and 6B) [96]. Experiments show that the results measured by LM coil accords well with that by classical copper coil during eye movement. Meanwhile, Fig. 6C has fully showcased the excellent flexibility of LM coil. Coupled with its flexibility, these coils show great advantages in implantable system, such as retinal prosthesis system, whose secondary coil size is strictly limited by the anatomy of eye. Based on these studies, LM-based coils may hopefully be applied to wireless battery chargers, cardiac implants and glucose monitoring implants.

Stefan et al. had ever designed a new cochlear implant electrode with LM [64]. Commonly, the cochlear implant system transfers auditory signals to brain by activating auditory nerves. Unlike typical cochlear implant systems, this design applies LM to electrode wires, with higher flexibility and biocompatibility. What's more, the array carrier used in this design is reactive to LM, which makes sure that a local leakage seal will form when a break occurs so that LM will not migrate to the outer surface. Another LM-based cochlear implant electrode combining with microfluidic is designed with good conductivity and flexibility to break the limitation of tradition electrodes (Fig. 6D). The electrode is highly thin and soft so that it does little physical damage to the internal structure of cochlea. As a result, the residual hearing can be protected.

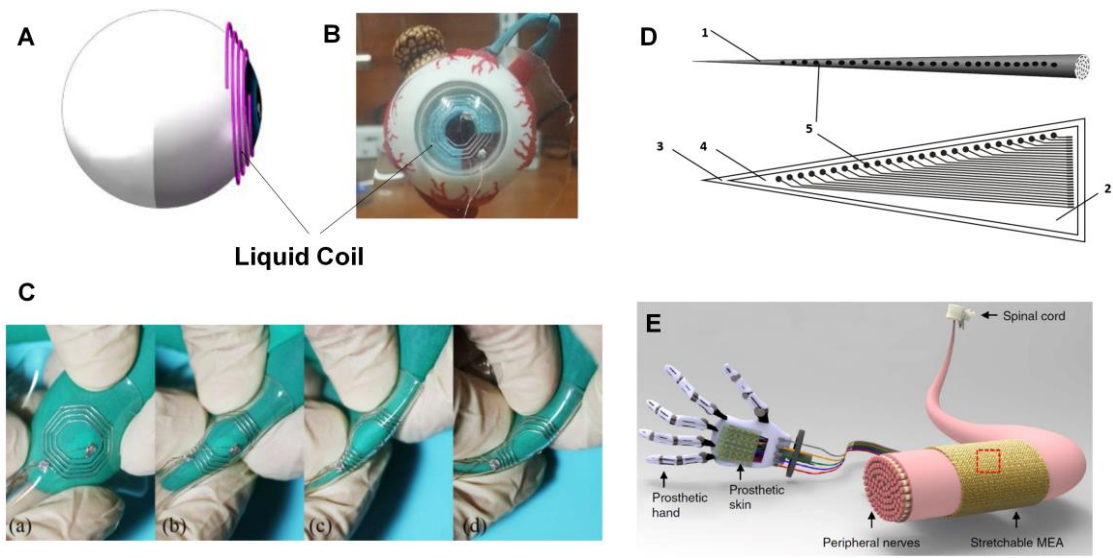

**Figure 6.** (A)The geometric model of LM induction coil. (B) The LM coil embedded in PDMS is put on the artificial eyeball for eye movement recording [96]. (C) Bending the LM coil to show its flexibility [96]; (D) Cochlear implant electrodes based on LM: (1) A cochlear implant electrode, (2) PDMS membrane, (3) Lower membrane including the microfluidic channels, (4) Upper membrane, (5) Complex of LM and silica gel. (E) A prosthetic hand [97].

**3.5 Prostheses**

Recent research has been conducted to apply LM to prostheses. Prostheses can be mainly divided into two types, aesthetic functional devices and myoelectric devices. Particularly, myoelectric prostheses work by sensing electrical signal from nerves and muscles via electrodes. Combining LM electrodes with signal process systems, the motion of prostheses can be controlled by users precisely. To meet even higher standards, a smart prosthetic hand should not only receive the instructs from nervous system and make corresponding responses, but also feel various types of external stimuli and transmit signals to the corresponding peripheral nervous system to realize better control of prostheses (Fig. 6E) [97, 98]. For this purpose, artificial sensing skin and myoelectric prostheses can be combined to makes the prostheses dexterous and smart.

With the development of microfluidic technologies, LM can be used to produce soft tactile sensors, which has great potential in sensing skin, such as temperature, sweat analysis, skin patch and pressure, based on the change of resistance and capacitance. For instance, Jung et al. combined microfluidic channel and LM together to make a pressure sensor [99]. Additionally, an inertial sensor has been designed based on the capacitance between two LM electrodes, which can measure tilt angles and record arm gestures [65]. Liu et al. have discovered a 3-D micro biological electrode array by LM micromolding technique [72]. Due to its miniaturization, the microneedle bio-electrode can effectively decrease the contact resistance with skin, and therefore it can be used to fabricate micro tactile sensors to detect weak signal changes. These designs in soft tactile sensors promote advances in artificial sensing skin and increase the possibility of successfully applying artificial sensing skin to prostheses. It can be concluded that with soft and stretchable LM sensors and electrodes, a comfortable and smart prosthesis for amputees to use is promising.

**3.6 Cardiac pacing**

For the treatment of heart-rhythm diseases, advanced capabilities in electrical recording and stimulating are essential [100]. Cardiac electrotherapies include pacing, defibrillation and cardiac ablation therapy [101]. Stretchable electronics show high performance in cardiac electrotherapies with the development of new materials.

Combining with specific sensor or chip, multifunctional soft devices are produced. Attaching the device onto soft skin, stable physiological signals (such as EEG (electroencephalogram), ECG (electrocardiogram), sweat, temperature, pulse and gait etc.) are detected even in human activities. Intelligent electronic skin network (Fig. 7A) based on LM flexible electronics is expected to come true in the future with the functions of sensing, monitoring and treating [102]. Besides, Jin et al. proposed the original idea of 3D injectable electronics and injected LM electrode into the left side of the thorax near the left upper limb of a mouse. Then electrical stimulation with different magnitude (0.6 mV, 1.2 mV and 1.3mV) was applied to the mouse and ECG signal was recorded [59]. When magnitude increased, PQ interval of ECG signal shortened and heart rate increased (Fig. 7C). The phenomenon that the increase of electrical stimulation induces the spasm of mouse muscle inspires the applications of implanted electrical stimulation on heart pace-making and defibrillation. ECG signals are initiated by a pulse of electrical excitation from the specialized pacemaker cells, which make the heart beat rhythmically. The repolarization of the heart cell (Fig. 7D site 1) is interrupted by depolarization (Fig. 7D site 2), enabling the electrical impulse to spread from cell to cell in the syncytial heart [51]. Additionally, stretchable wireless power transfer device can be prepared by LM circuits with excellent flexibility and stable energy conversion [103]. All these results provide possibility that LM can be applied to recoding cardiac activity and delivering electrical stimulation to treat cardiac diseases (Fig. 7E) [101].

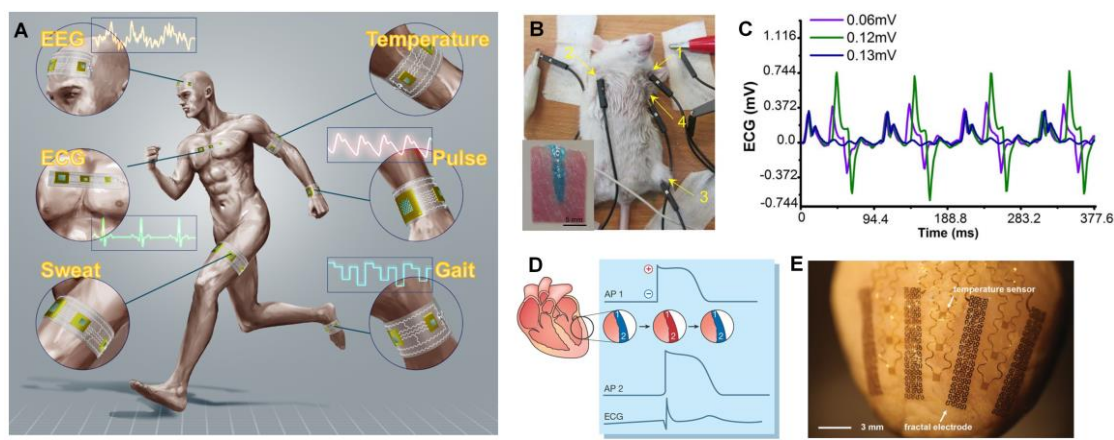

**Figure 7**. (A) The intelligent electronic skin network based on LM e-skin and wearable bioelectronics [102]. (B) Illustration for conducting an electrical stimulation to the experimental mouse [59]. (C) The recorded ECG signals of experimental mouse undergoing a 10 Hz electrical stimulation with magnitude of 0.6 mV, 1.2 mV and 1.3 mV, respectively [59]. (D) The formation of ECG signals; Negatively polarized (resting) muscle is shown as blue; depolarized muscle is red; Action potentials recorded at sites 1 and 2 are similar in timing and morphology [51]. (E) Image of a representative device integrated on a Langendorff-perfused rabbit heart. The white arrows highlight functional components [101].

### 3.7 Tumor treating

Some *in vitro* evidences have demonstrated the efficacy of tumor treating electrical fields on cancer therapy, such as human melanoma, glioma, lung, prostate, pancreas, and breast; mouse adenocarcinoma and melanoma; and rat glioma [48, 104, 105]. From Section 2, we can draw the conclusion that electrostimulation in cancer therapy mainly inhibits and kills tumor cells through two mechanisms: disrupting cell mitosis, and increasing cell apoptosis. Li et al. sprays LM on the malignant melanoma tumor on C57BL/6 mice and adds a sine wave electrical power to deliver electrical stimulation to tumor tissue [71]. After one-week treatment, the tumors with treating by 2 V/cm, 300 kHz and 1.6 V/cm, 300 kHz electrical stimulation diminish (Fig. 8E). Unlike Li et al.'s spray-printing method, Sun et al. put forward a new method, injection. Though injecting, the shape of LM electrodes is easily changed to achieve better performance of tumor treating (Fig. 8A). Comparing *in vitro* cell viability and *in vivo* antitumor effects of conformable LM electrodes and conventional platinum (Pt) electrodes, LM electrodes provide not only transformable capabilities, but also a highly efficient electron-transfer ability, which allows better tumor destruction effects in both anode and cathode areas [12]. It is hopeful that LM electrodes can help to regulate various desirable configurations, two or even three dimensions, which can be applied to irregular tumor surface. What's more, the electrical field in tumor treatment can destroy tumor cells with no appreciable heat effect or stimulation on surrounding muscles and nerves significantly. Different from rigid electrodes or sol-gel electrode materials, the LM sprayed on skin provides biggest surface area, which diminishes the contact resistance and therefore reduces the burning rate of electrical stimulation sites.

In addition, Fig. 8C displays the microscopic evidence of HeLa cells membrane blebbing with electrical field treating during mitosis [24], which reveals violent structural disruption. The promising results offer evidence to electrical stimulation by LM electrodes for further cancer therapy.

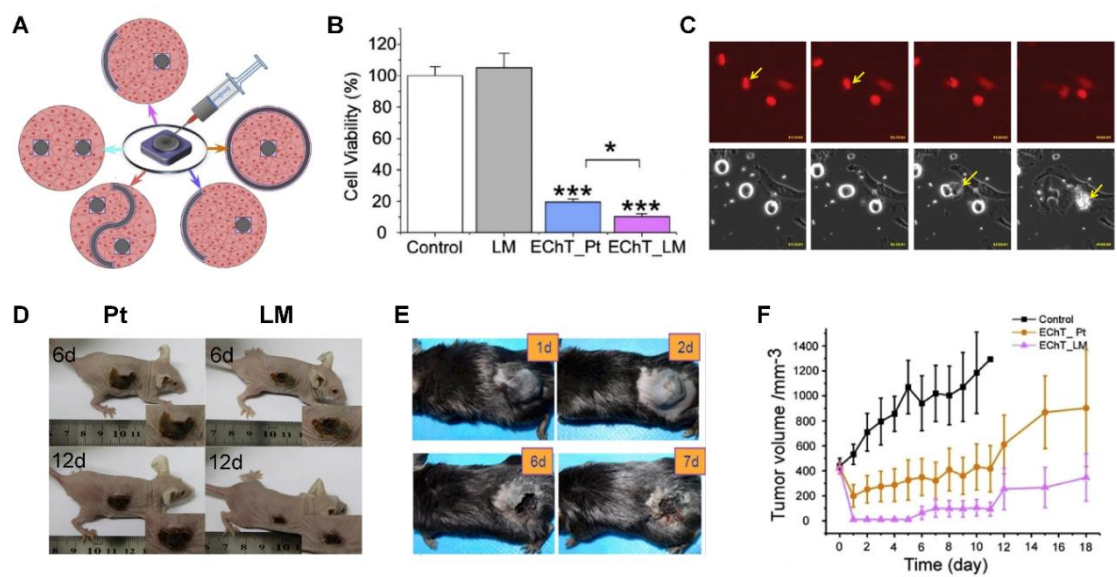

**Figure 8.** Electrostimulation in tumor treatment. (A) Five therapeutic method configurations of LM electrode in different shapes [12]; (B) Cell viability of LM, EchT_Pt, EchT_LM [12]; (C) The microscopic evidence of HeLa cells membrane blebbing with TTFields treatment during mitosis [24]; (D) The 6$^{th}$ and 12$^{th}$ days treatments of melanin tumor-bearing mice by Pt electrode and LM electrode treating [12]; (E) Electrical stimulation treatment of malignant melanoma tumor on C57BL/6 mice based on LM spray-printing [71]; (F) Tumor volume of different groups of tumor-bearing mice [12].

### 3.8 Immunotherapy and psychotherapy with LM bath electrode

Immunization therapy in the present paper refers to the enhancing or inhibiting of organism immune system due to introduction of electrical stimulation to achieve desired therapeutic output. Unlike conventional immunization therapy through injecting immunostimulant or immunosuppressive drugs into patient's body, the novel immunization therapy will enable minimally invasive disease therapy with no drug to inject into human body and applying LM bath electrode to conformably and electrically stimulate organism [69]. Besides, when electricity stimulates human, we will have lots of strange feelings. With regard to virtual reality, LM enabled electrobiology can create a virtual environment and regulate human's emotions, and therefore be used in psychotherapy. Imaging the future life, patient can enjoy his treatment whatever physical or mental diseases in the LM bath with specific amplitude and frequency electricity stimulation.

Overall, considering all the above potential routes to tackle modern disease challenges, we can summarize the generalized way of administering LM enabled electrobiology as Fig. 9. Clearly, there is plenty of space to explore in the coming time along this direction.

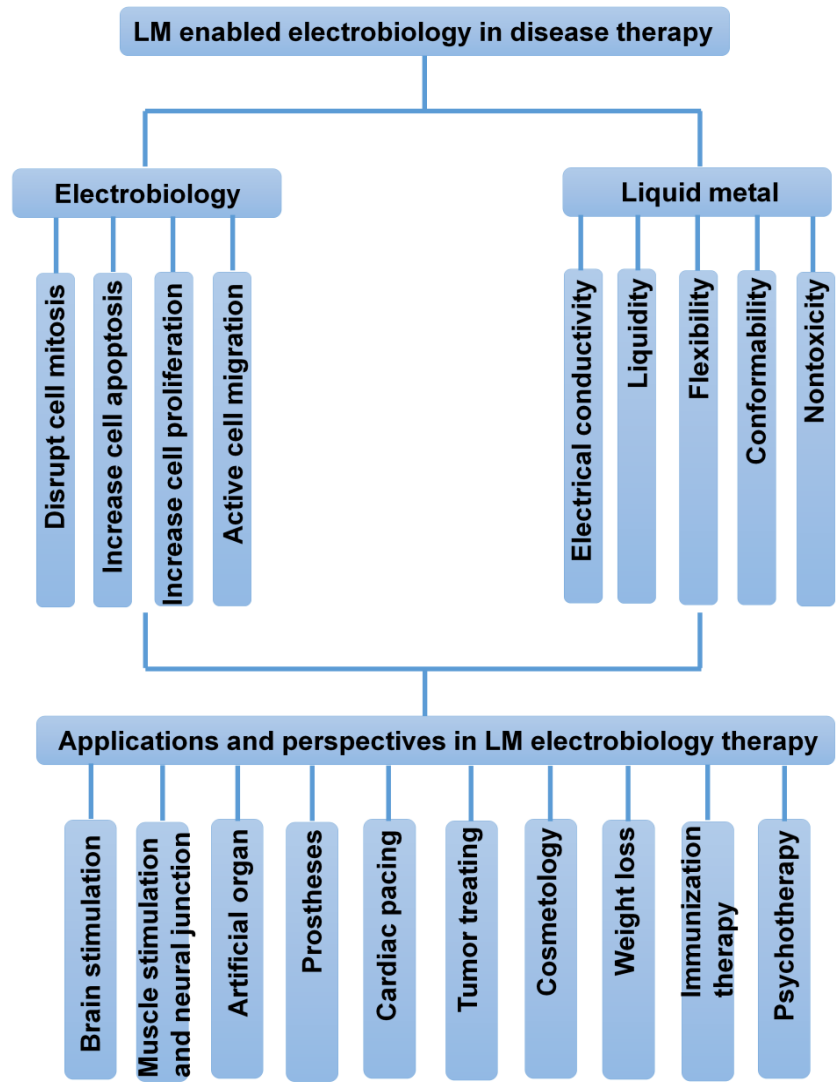

**Figure 9.** The summary diagram of LM enabled electrobiology in disease therapy

## 4. Conclusion

In summary, a basic concept of LM electrobiology is proposed and a generalized disease therapy framework is established for the coming studies and clinical clarifications. Due to its remarkable material features, such as metallicity, reliable flexibility, excellent electrical conductivity and non-toxicity, LM can not only well adapt to the body disease models to generate conformable electrical treatment, but also reduce the contact resistance to avoid more energy loss when performing the task. At this stage, several basic electrostimulation mechanisms in disease therapy have been investigated. One of them cures diseases by inducing cell proliferation, migration and excitation, including electrical muscle stimulation, nerve connection, artificial organs, cardiac pacing etc. The others prevent the cell lines by disrupting cell mitosis or increasing cell apoptosis. Taking oncotherapy as an example, under specific tumor treating field for a period of time, the tumor diminishes along with cancer cells being killed. Overall, given specific design, typical applications of LM electrostimulation would provide satisfactory therapeutic efficacy, conformable treatment area and biological availability. This may lead to an ever easy-going physical way for the coming clinics and health care.

## Acknowledgements

This work is partially supported by the Ministry of Higher Education Equipment Development Fund, Tsinghua University Fund, NSFC Key Project under Grant No. 91748206, Dean's Research Funding and the Frontier Project of the Chinese Academy of Sciences.